\documentclass[showpacs,aps,epsfig,nofootinbib]{revtex4}
\usepackage{color,soul}
\usepackage{graphicx}
\usepackage{amsmath}
\usepackage{amsfonts}
\usepackage{graphicx}
\usepackage{epstopdf}
\usepackage{latexsym}
\usepackage{amssymb}
\usepackage{color}
\usepackage[brazilian]{babel}
\usepackage[utf8]{inputenc}
\usepackage[T1]{fontenc}
\usepackage[center]{subfigure}

\begin{document}

%%%%%%%%%%%%%%%%%%%%%%%%%%%%%%%%%%%%%%%%%%%%%%%%%%%%%%%%%%%%%%%
 \newcommand{\bq}{\begin{equation}}
 \newcommand{\eq}{\end{equation}}
 \newcommand{\bqn}{\begin{eqnarray}}
 \newcommand{\eqn}{\end{eqnarray}}
 \newcommand{\nb}{\nonumber}
 \newcommand{\lb}{\label}
\newcommand{\PRL}{Phys. Rev. Lett.}
\newcommand{\PL}{Phys. Lett.}
\newcommand{\PR}{Phys. Rev.}
\newcommand{\CQG}{Class. Quantum Grav.}
%
 %%%%%%%%%%%%%%%%%%%%%%%%%%%%%%%%%%%%%%%%%%%%%%%%%%%%%%%%%%%%%%%

\title{(Anti-) de Sitter Electrically Charged Black Hole Solutions in Higher-Derivative Gravity}

\author{Kai Lin$^{1)}$}
\email{lk314159@hotmail.com}

\author{Wei-Liang Qian$^{2,3)}$}
\email{wlqian@usp.br}

\author{A. B. Pavan$^{1)}$}
\email{alan@unifei.edu.br}

\author{E. Abdalla$^{4)}$}
\email{eabdalla@usp.br}

 \affiliation{$^{1)}$Instituto de F\'isica e Qu\'imica, Universidade Federal de Itajub\'a, MG, Brasil}
 \affiliation{$^{2)}$Escola de Engenharia de Lorena, Universidade de S\~ao Paulo, SP, Brasil}
 \affiliation{$^{3)}$Faculadade de Engenharia de Guaratinguet\'a, Universidade Estadual Paulista, SP Brasil}
 \affiliation{$^{4)}$Instituto de F\'isica, Universidade de S\~ao Paulo, CP 66318, 05315-970, S\~ao Paulo, Brasil}

\date{\today}

\begin{abstract}
In this paper, static electrically charged black hole solutions with cosmological constant are investigated in an
Einstein-Hilbert theory of gravity with additional quadratic curvature terms.
Beside the analytic Schwarzschild (Anti-) de Sitter solutions, non-Schwarzschild (Anti-) de Sitter solutions are
also obtained numerically by employing the shooting method.
The results show that there exist two groups of asymptotically (Anti-) de Sitter spacetimes for both charged and
uncharged black holes. In particular, it was found that for uncharged black holes the first group can be reduced
to the Schwarzschild (Anti-) de Sitter solution, while the second group is intrinsically different from a
Schwarzschild (Anti-) de Sitter solution even when the charge and the cosmological constant become zero.
\end{abstract}

\pacs{04.70.Bw, 04.25.dg, 04.60.-m}

\maketitle

Though Einstein's general relativity has been extensively tested at the highest achievable experimental precision
up to date, gravity is not a renormalizable quantum field theory from the theoretical viewpoint.
A possible attempt to solve the problem of the non-renormalizability of general relativity is to
include higher-order corrections that become important at higher energy \cite{tHooft}.
In this context, general relativity can be viewed as an effective low-energy field theory, which is
understood to be largely independent of the details in higher energy scales. It can be shown when all
possible quadratic curvature invariants are added to the Einstein-Hilbert action, a renormalizable theory is
achieved at the cost of the presence of the ghost modes \cite{stelle}. Thus the study of the properties of
such higher-derivative gravity could shed light on the ongoing efforts to understand the nature of gravity.

In order to better understand such a new theory of gravity, it seems of considerable interest to investigate the
behaviour  of black hole solutions, since {they} are fundamental objects in general relativity.
Such studies have also been performed in other theories of gravity containing higher order curvature corrections with well defined
coefficients, such as Lovelock theory \cite{lovelockkaiabdoliv}.
Moreover, the question of stability of charged Black Holes is relevant and has been intensively studied recently in recent years
\cite{pellicerwangabdalla,konoplyazhidenko}.

Recently, L\"u \emph{et. al} found, numerically, a non-Schwarzschild solution in a theory of gravity with a quadratic
Weyl scalar on action \cite{LPPS,LPPS1}, and revealed that this higher-derivative gravity possesses {additional}
static spherically symmetric black hole solution {comparing} to Einstein's general relativity.
{We also showed} in \cite{LPFA} that electrically charged static solutions in this theory are also characterized by two
groups of solutions. These solutions reduce respectively to Schwarzschild black hole and non-Schwarzschild black
hole solutions, when the charge of the black hole goes to zero. Surprisingly the Reissner-Nordstr\"om metric is
not a solution of the field equations when the coefficients of the higher-derivative terms do not vanish.

Being first introduced by Einstein to describe a static universe and later abandoned by himself after Hubble observes
the expansion of universe, the cosmological constant is a measure of the value of the energy density of the vacuum
of space and could be a candidate to explain the expansion. In addition, the discovery of the accelerating universe
after 1990 from distant supernovae implies that more than half of the energy density of the universe can be
attributed to the unkown so called dark energy.
Thus, while poorly understood at a fundamental level, the concept of cosmological constant was revived and is,
by far, the simplest possible form of dark energy.
Another important motivation in studying spacetimes with cosmological constant, namely negative one, is the
AdS/CFT correspondence \cite{adscft}.
Therefore, it seems to be interesting to investigate the properties of the non-Schwarzschild black hole solutions
with the presence of the cosmological constant.
In this letter, we carry out a study of the (Anti-) de Sitter electrically charged black hole solutions in
higher-derivative gravity.

The action of higher-derivative gravity in Einstein-Hilbert theory with quadratic curvature scalars and with
electromagnetic field and the cosmological constant $\Lambda$ can be written as
 \bq
\label{eq1} {\cal L}=\gamma R-2\Lambda-\alpha
C_{\mu\nu\rho\sigma}C^{\mu\nu\rho\sigma}+\beta R^2-\kappa
F_{\mu\nu}F^{\mu\nu},
 \eq
where $F_{\mu\nu}=\nabla_\mu A_\nu-\nabla_\nu A_\mu$ is the electromagnetic tensor, $C_{\mu\nu\rho\sigma}$ is the
Weyl tensor, $\alpha$, $\beta$, $\gamma$ and $\kappa$ are coupling constants.

According to the analysis in \cite{LPPS,LPFA,WN}, (non-)charged black hole solutions without cosmological
constant are independent of the $\beta R^2$ term since the additional terms on the action are traceless. Here,
this condition is broken because of the introduction of the cosmological constant. However, we set $\beta=0$
for the sake of convenience. Thus, the resulting field equations are given by

\bqn
 \label{eq2a}
R_{\mu\nu}-\frac{1}{2}g_{\mu\nu}R-\Lambda g_{\mu\nu}-4\alpha B_{\mu\nu}-2\kappa T_{\mu\nu}&=&0,\\
 \nb\\
 \label{eq2b}
 \nabla_{\mu}F^{\mu\nu}&=&0.
 \eqn
where the trace-free Bach tensor $B_{\mu\nu}$ and energy-momentum tensor of electromagnetic field
$T_{\mu\nu}$ are defined as
 \bqn
\label{eq3a}
B_{\mu\nu}&=&\left(\nabla^\rho\nabla^\sigma+\frac{1}{2}R^{\rho\sigma}\right)C_{\mu\rho\nu\sigma},\\
 \nb\\
 \label{eq3b}
T_{\mu\nu}&=&F_{\alpha\mu}F^{\alpha}_\nu-\frac{1}{4}g_{\mu\nu}F_{\alpha\beta}F^{\alpha\beta}.
 \eqn
For a static black hole with spherical symmetry, the metric has the form
 \bq
\label{eq4}
ds^2=-h(r)dt^2+\frac{dr^2}{f(r)}+r^2d\theta^2+r^2\sin^2(\theta)d\varphi^2\quad .
 \eq
By substituting Eq.(\ref{eq4}) into the field equations, we arrive at the differential equations
 \bqn
\label{eq5}
 r h \left[r f' h'+2 f\left(r h''+2h'\right)\right]+4h^2 \left(r f'+f-1+2\Lambda r^2\right)-r^2 f h'^2&=&0\\
 \nb\\
 \label{eq5a}
 f''+\frac{r^2 fh'^2+2 r f h h'+4 (f-1+2\Lambda r^2)h^2}{2 r f h \left(r h'-2 h\right)}f'-\frac{3 h f'^2}{4 f h-2 r f
 h'}+8\Lambda\frac{h-f h-r f h'-\Lambda r^2h}{3f(2h-rh')}&&\nb\\
 \nb\\
 -\frac{r^3f h'+\left(r^2f-r^2+\Lambda r^4+\kappa Q_0^2\right)h}{\alpha r^2 f \left(r h'-2 h\right)}
-\frac{r^3fh'^3-3r^2fhh'^2-8(f-1)h^3}{2r^2 h^2 \left(r h'-2
 h\right)}&=&0\\
 \nb\\
 \label{eq5b}
 A_t'+\sqrt{\frac{h}{f}}\frac{Q_0}{r^2}&=&0
 \eqn
where $Q_0$ is the electric charge of the black hole. Imposing standard physical boundary conditions, implies
that $A_t(r)\to 0$ at the cosmological horizon $r_c$ (infinity) in an asymptotically (Anti-) de Sitter spacetime.
Besides, it is easy to verify that, for uncharged spacetimes, the analytic Schwarzschild
(Anti-) de Sitter black hole solutions satisfy the above field equations.
However, this is not true for charged spacetimes. It can be shown straightforwardly that the Reissner-Nordstr\"om
(Anti-) de Sitter metric does not satisfy the above field equations when $\alpha\not=0$. The same condition was
pointed out in \cite{LPFA} for the case without cosmological constant.

In what follows we apply the shooting method to numerically study the Schwarzschild as well as non-Schwarzschild
solutions and their generalizations. We can set $\alpha=\frac12$ and $\kappa=1$ without losing generality. After
that, we expand the functions $h(r)$ and $f(r)$ around the event horizon $r_0$ as follows,
 \bqn
\label{eq6}
 h(r)&=&h_1(r-r_0)+h_2(r-r_0)^2+h_3(r-r_0)^3+\cdot\cdot\cdot,\nb\\
 f(r)&=&f_1(r-r_0)+f_2(r-r_0)^2+f_3(r-r_0)^3+\cdot\cdot\cdot,
 \eqn
where $f_i$ and $h_i$ are constant coefficient. Since we can always rescale the time coordinate by the transformation
$t\rightarrow\frac{t}{\text{Constant}}$ for any positive constant $C$, then $C\times h(r)$ is always a solution of
the field equations if $h(r)$ itself is a solution of the corresponding equations. Therefore, we may choose
$f_1=h_1$ for the sake of simplicity. By substituting Eq.(\ref{eq6}) into the field equations, all $h_j$ and
$f_j$ with $j\ge 2$ can be expressed in terms of $f_1$, for example, $h_2$ and $f_2$ are given by
 \bqn
\label{eq7}
 h_2&=&\frac{1-2f_1r_0}{r_0^2}+\frac{r_0^2-f_1r_0^3-\kappa Q_0^2}{8\alpha f_1r_0^3}-\left(\frac{5}{3}+
\frac{1}{3f_1r_0}+\frac{3r_0}{8\alpha f_1}\right)\Lambda+\frac{r_0\Lambda^2}{3f_1},\nb\\
 f_2&=&\frac{1-2f_1r_0}{r_0^2}-3\frac{r_0^2-f_1r_0^3-\kappa Q_0^2}{8\alpha f_1r_0^3}-\left(3-\frac{1}
{f_1r_0}-\frac{3r_0}{8\alpha f_1}\right)\Lambda-\frac{r_0\Lambda^2}{f_1}.
 \eqn

Now we are in the position to calculate the numerical black hole solutions of the field equations.

For the case of asymptotically de Sitter black holes ($\Lambda>0$) we fix $\Lambda=0.01$ for the sake of convenience.
Let us first study the case when $Q_0=0$, which corresponds to the uncharged black holes. As mentioned above, the
Schwarzschild de Sitter metric is an exact solution of the field equations. The non-Schwarzschild de Sitter
solutions were found numerically by the following procedure. Since the de Sitter black hole has an event horizon
and a cosmological horizon, $f$ and $h$ should satisfy the boundary conditions to attain zero at both horizons.
From the form of Eq.(\ref{eq6}), it is obvious that the above first boundary condition at the event horizon is
automatically satisfied, and the second boundary condition at cosmological horizon is used as the
criterion of the initial value of $f_1$ in the shooting method.

\begin{figure}[h]
\includegraphics[width=6cm]{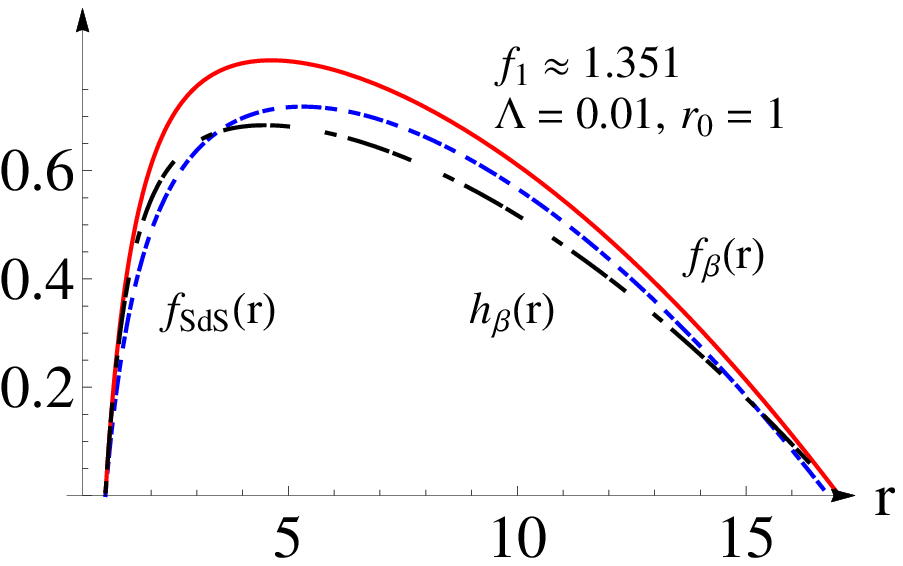}\includegraphics[width=6cm]{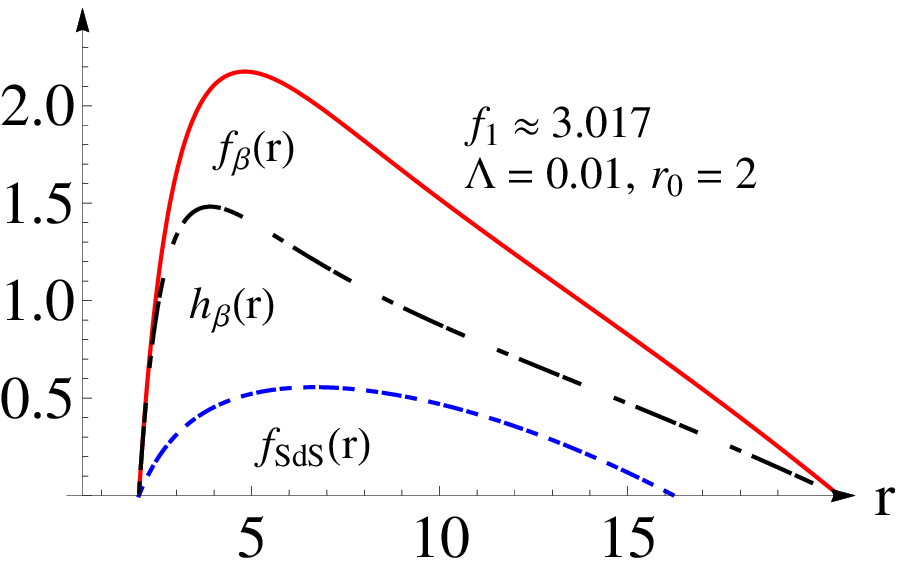}\includegraphics[width=6cm]{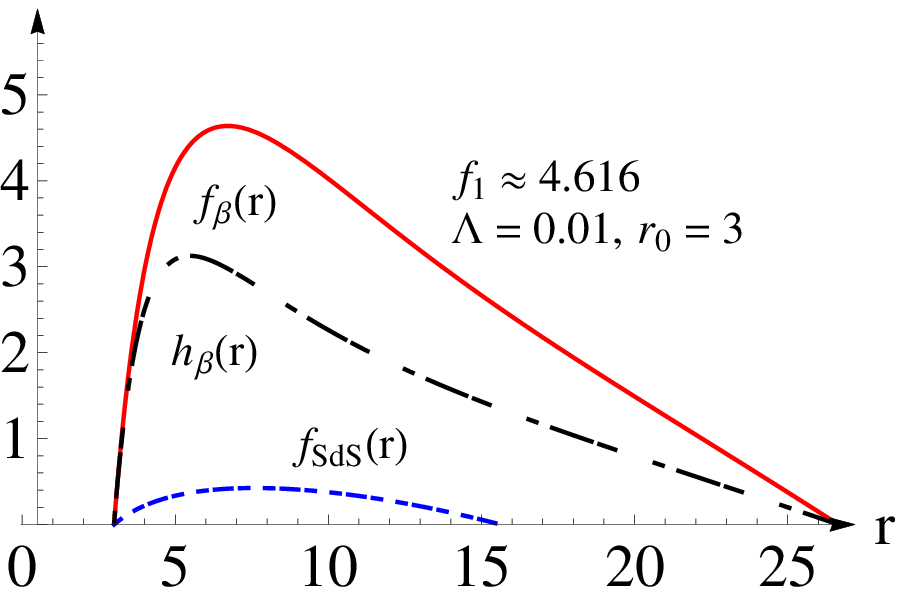}
\caption{Schwarzschild de Sitter black hole and non-Schwarzschild de Sitter black hole for $r_0=1$, $r_0=2$ and $r_0=3$.
We have $f(r)=h(r)=f_{SdS}(r)$ for the Schwarzschild de Sitter case, and $f(r)=f_\beta(r)$ and $h(r)=h_\beta(r)$ for
the non-Schwarzschild de Sitter case. The group I solution (left) and group II (middle, right)} \label{fig2}
\end{figure}

At this point we shall introduce a classification to the black hole solutions that will be used throughout paper.
The Schwarzschild (Anti-) de Sitter solutions and those that can be reduced to the Schwarzschild (Anti-) de
Sitter solutions under specific conditions are referred to as group I solution. Similarly, the non-Schwarzschild
(Anti-) de Sitter solutions as well as those can be reduced to the non-Schwarzschild (Anti-) de Sitter solutions
under specific conditions are referred to as group II.

\begin{figure}[h]
\includegraphics[width=5.5cm]{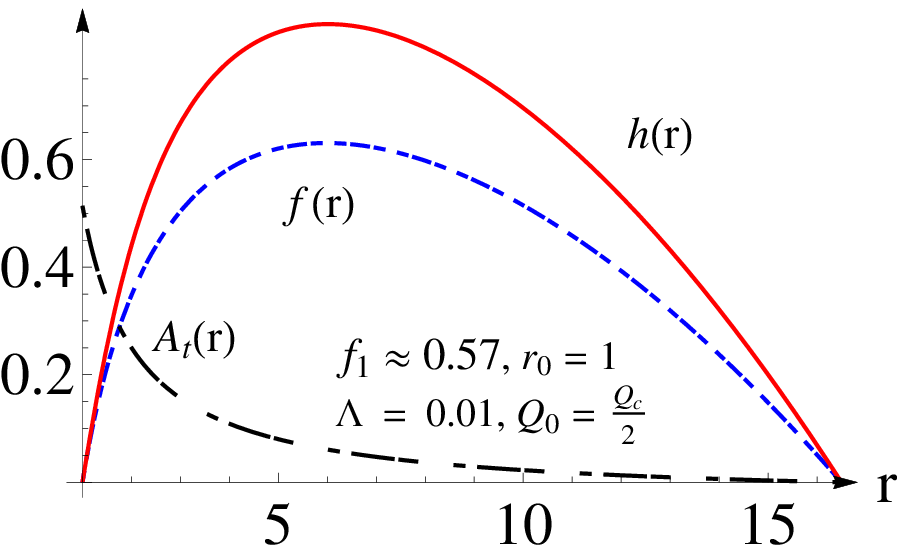}\includegraphics[width=5.5cm]{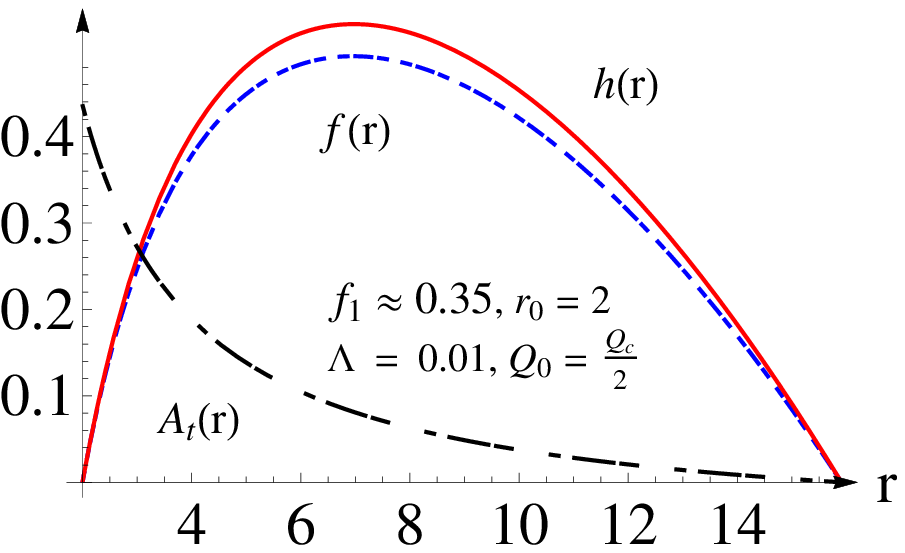}
\includegraphics[width=5.5cm]{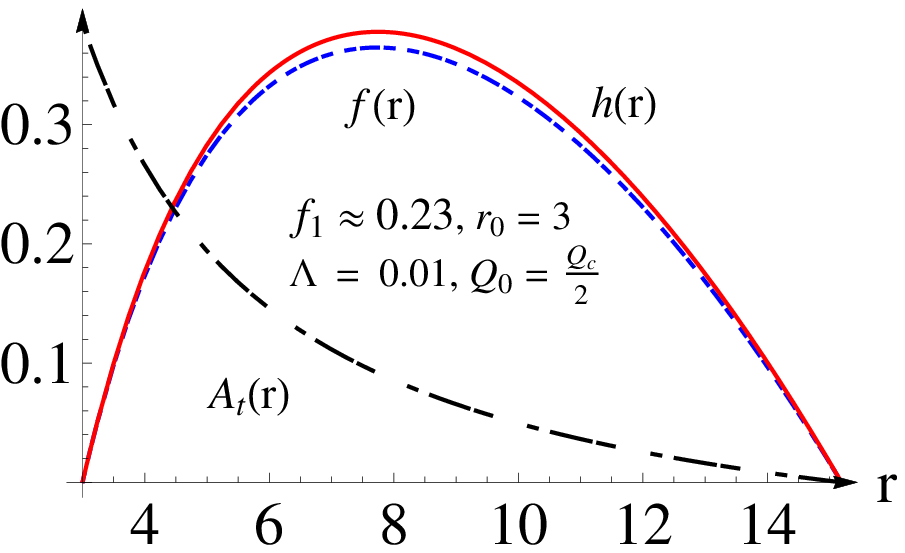}
\caption{Numerical charged de Sitter solutions $f(r)$, $h(r)$ and $A_t(r)$ belonging to group I with
$r_0=1,2,3$ and $Q_0=\frac{Q_c}{2}$.} \label{fig5}
\end{figure}

\begin{figure}[h]
 \includegraphics[width=5.5cm]{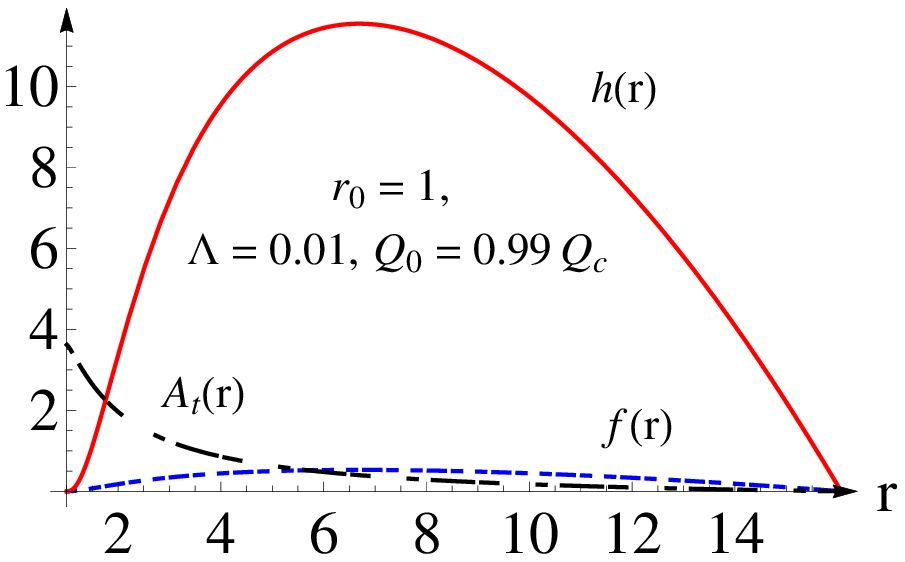}\includegraphics[width=5.5cm]{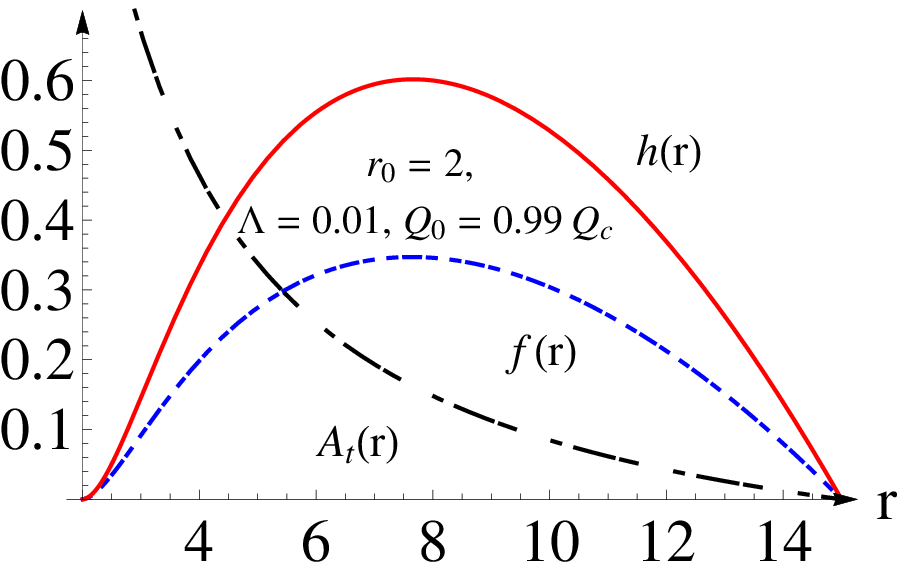}\includegraphics[width=5.5cm]
{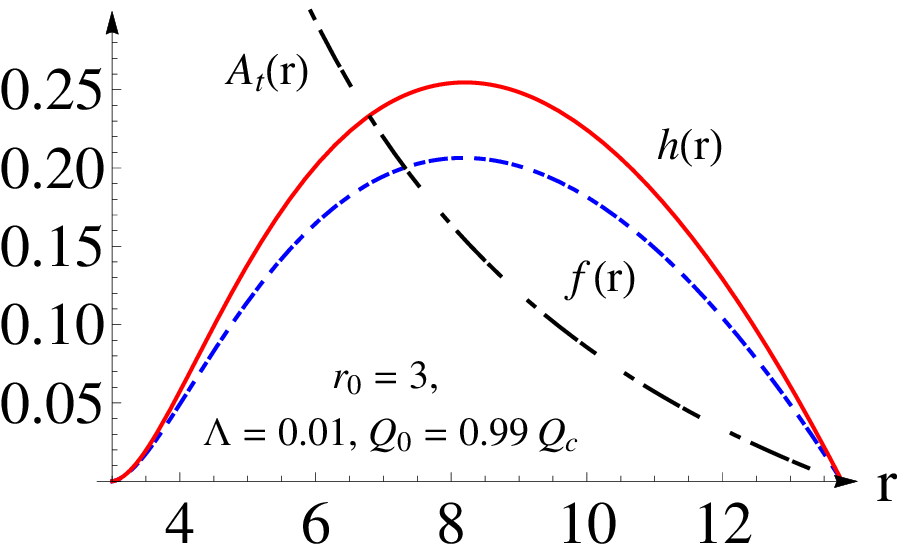} \caption{Numerical charged de Sitter solutions group I (near the extreme case) of $f(r)$, $h(r)$ and $A_t(r)$ with $r_0=1,2,3$
and $Q_0=0.99Q_c$, where $Q_c$ is the charge of the extreme de Sitter black hole (see Eq.(\ref{eq8})).} \label{fig5a}
 \end{figure}

\begin{figure}[h]
\includegraphics[width=5.5cm]{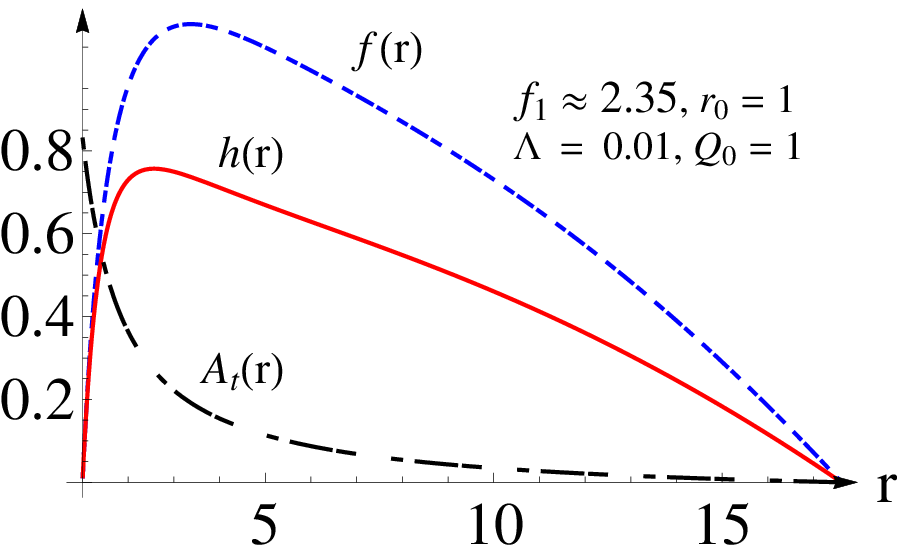}
\includegraphics[width=5.5cm]{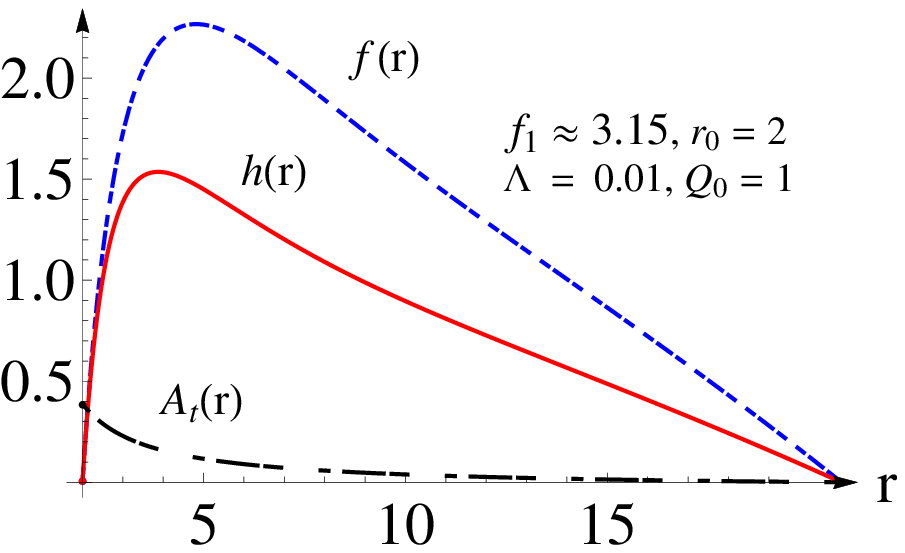}\includegraphics[width=5.5cm]{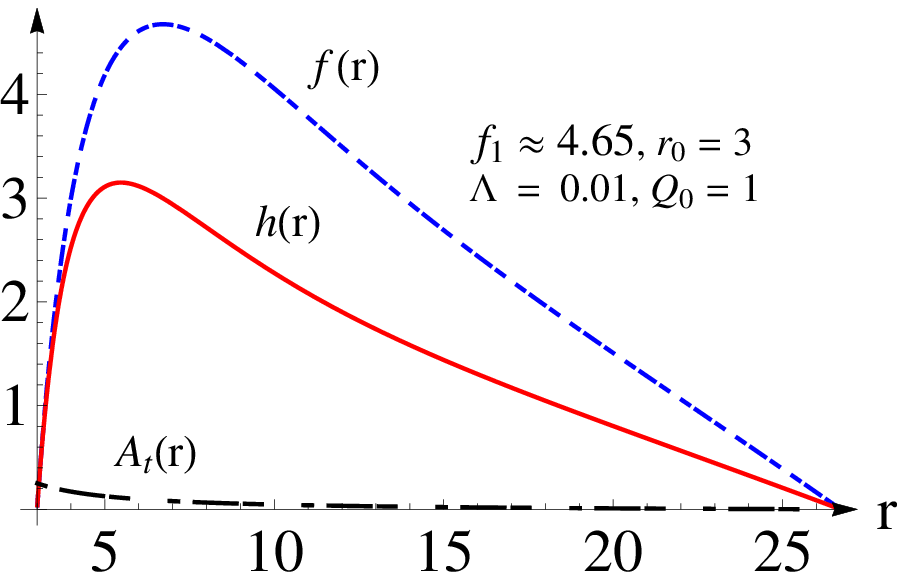}
\includegraphics[width=5.5cm]{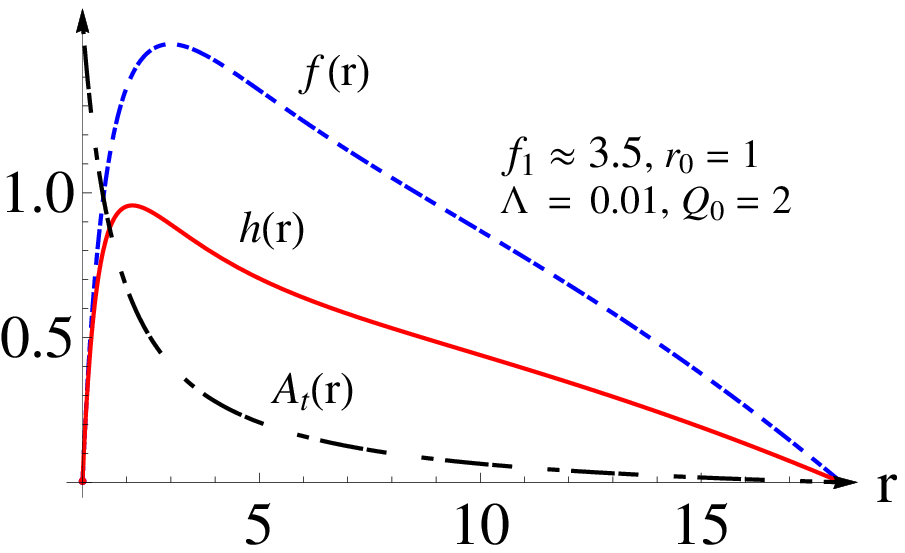}
\includegraphics[width=5.5cm]{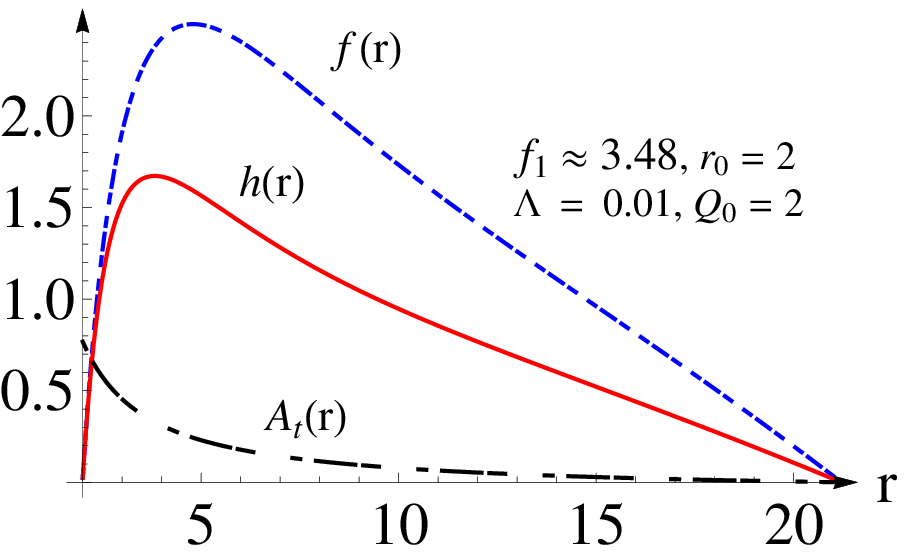}\includegraphics[width=5.5cm]{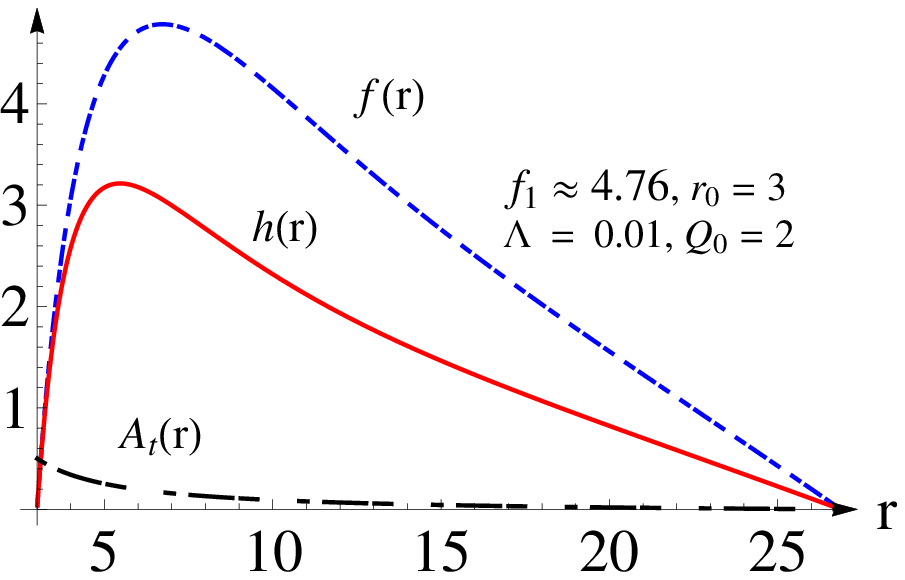}
\includegraphics[width=5.5cm]{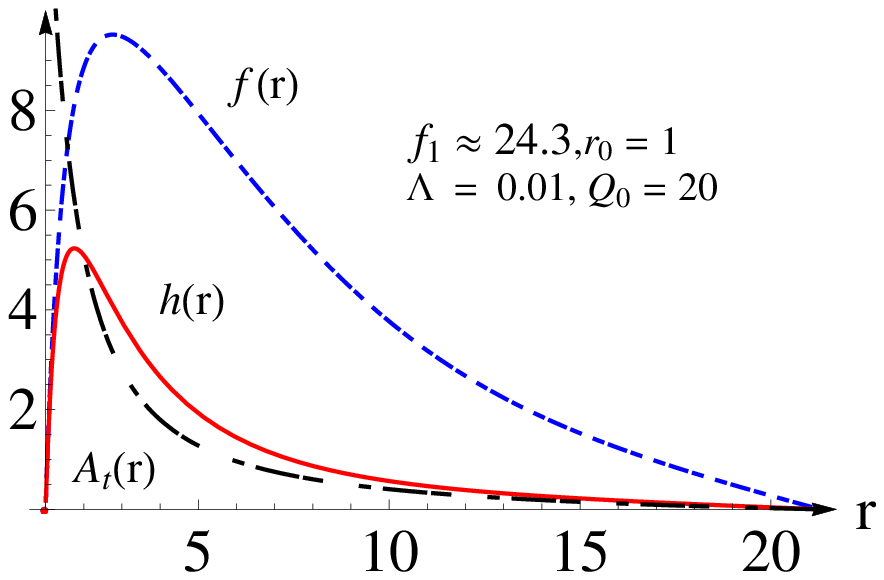}
\includegraphics[width=5.5cm]{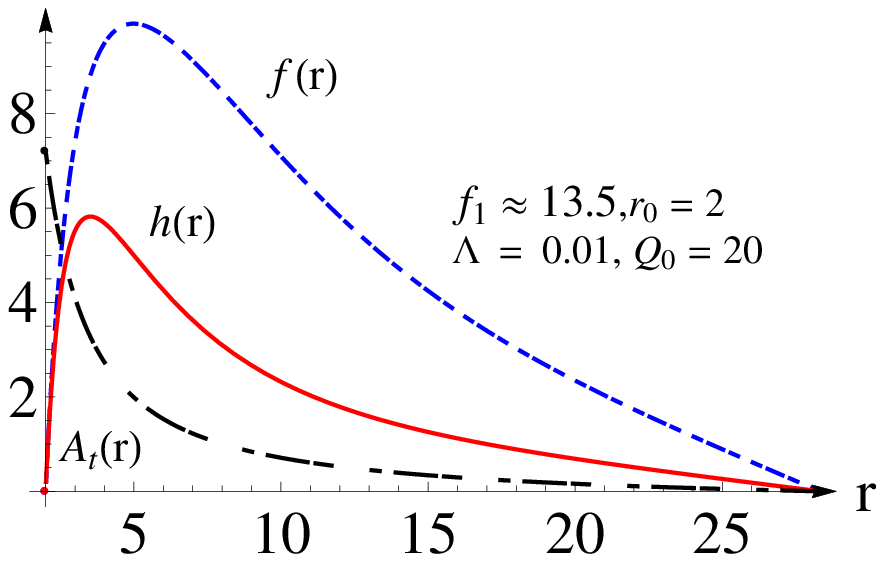}\includegraphics[width=5.5cm]{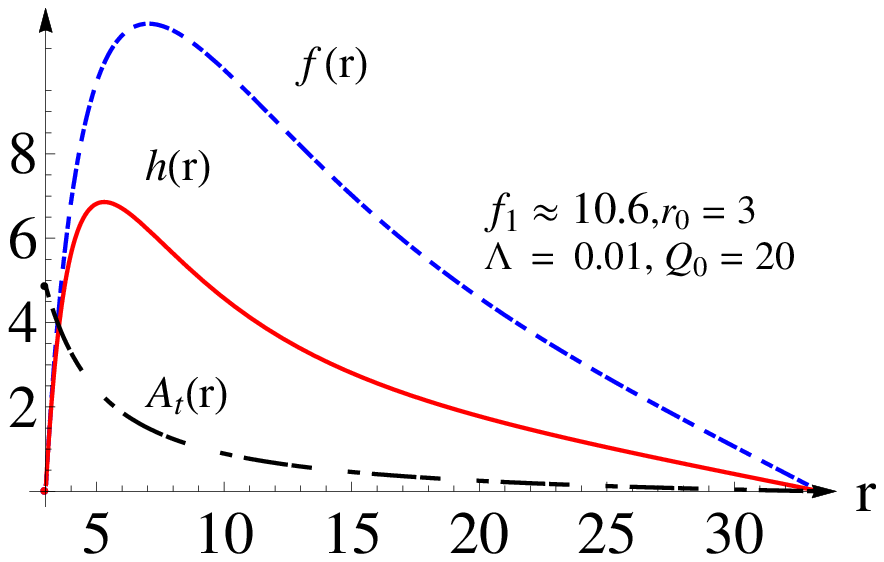}
\caption{Charged de Sitter solutions belonging to group II. Displayed are
$f(r)$, $h(r)$ and $A_t(r)$ with $r_0=1,2,3$ and $Q_0=1,2,20$.} \label{fig6}
\end{figure}

The (non-)charged asymptotically de Sitter black hole solutions found are shown in Figs
(\ref{fig2},\ref{fig5},\ref{fig5a},\ref{fig6}) for some given values of $r_0$ and $Q_0$.

When $Q_0=0$ and $\alpha\not=0$, we have two groups of solutions, such that group I reduce to Schwarzschild de
Sitter black hole and group II reduce to non-Schwarzschild de Sitter black hole respectively.
In Fig.(\ref{fig2}), one finds that the difference between Schwarzschild de Sitter black hole and non-Schwarzschild
de Sitter black hole increases as $r_0$ increases.  In special, the cosmological horizons are larger for
non-Schwarzschild de Sitter black hole than Schwarzschild de Sitter black hole. Besides, it is noted that
$h\not=f$ for non-Schwarzschild de Sitter black holes.

When $Q_0\not=0$ and $\alpha\not=0$, we can also separate the solutions in two groups, which reduce to Schwarzschild
de Sitter black hole and non-Schwarzschild de Sitter black hole respectively as $Q_0 \rightarrow 0$. In
Fig.(\ref{fig5}) are presented some charged asymptotically de Sitter solutions for different values of
$r_0$ and $Q_0=\frac{Q_c}{2}$.

For group I we found that there is a maximal physical value of the charge which is attained when the black hole
becomes an extreme de Sitter black hole. This critical value is given by
 \bqn
\label{eq8}
Q_c^{dS}=\frac{r_0}{\sqrt{3\kappa}}\sqrt{(r_0^2\Lambda-1)(8\alpha\Lambda-3)}\quad ,
 \eqn
and the extreme solution is reached imposing, additionally, $f_1=0$.
In  Fig.(\ref{fig5a}) we show some examples of the above results.
It is noteworthy that it is very difficult to directly calculate numerically the extreme case, so that we only study the case where the charge $Q_0$ is very close to the extreme value $Q_c$.
In the plots, we choose $Q_0=0.99Q_c$.
We note that the difference between $f(r)$ and $h(r)$ decreases with decreasing $r_0$.
In Fig.(\ref{fig6} we have the group II solutions. It turns out that the group II does not have an upper
limit of the charge differently of the group I. This interesting
propriety was already pointed out in \cite{LPFA} for the case
without cosmological constant. For both groups of solutions,
although they can be reduced to the (non-)Schwarzschild de Sitter
black hole at zero charge limit, none of them have
Reissner-Nordstr\"om de Sitter metric as valid solution. It can be
related with a non-trivial coupling between the electric charge and
the geometry through Weyl scalar and that appears in the
Eq.(\ref{eq7}).

\begin{figure}[h]
\includegraphics[width=6cm]{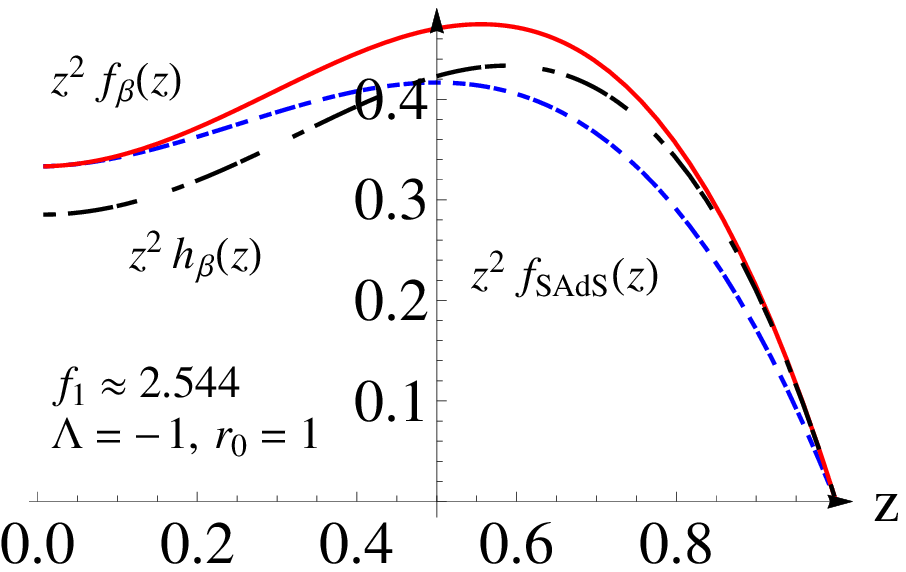}\includegraphics[width=6cm]{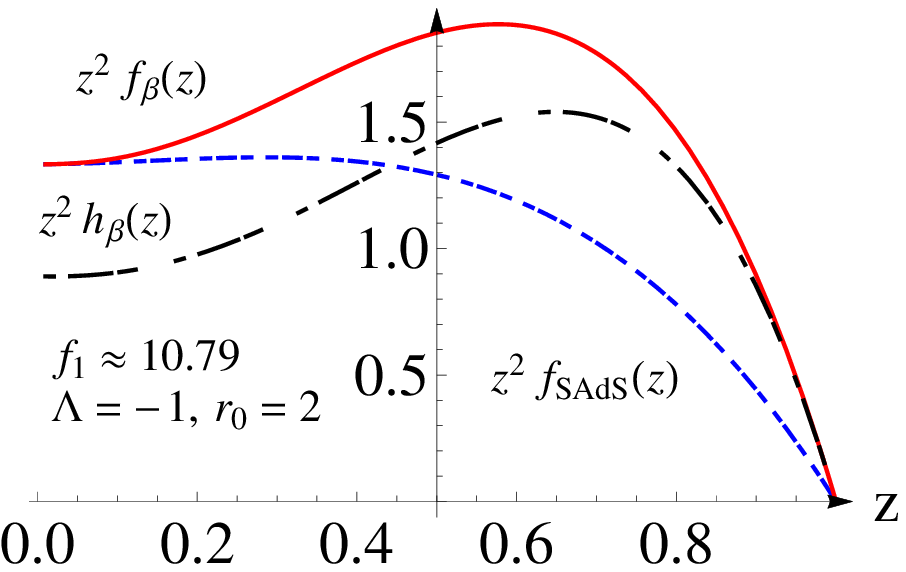}\includegraphics[width=6cm]{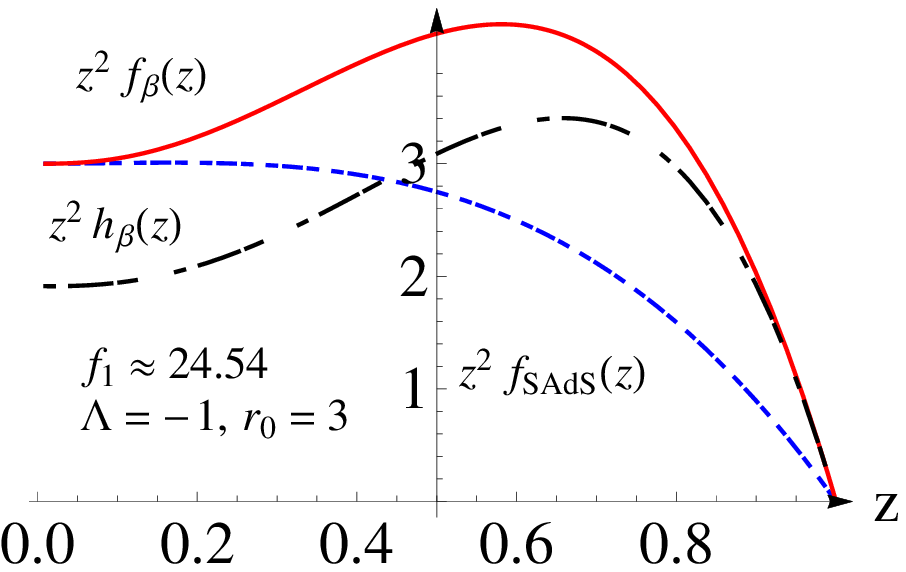}
\caption{Schwarzschild and non-Schwarzschild Anti-de Sitter black holes for $r_0=1$, $r_0=2$ and $r_0=3$.
We have $f(r)=h(r)=f_{ASdS}(r)$ for the Schwarzschild Anti-de Sitter case, and $f(r)=f_\beta(r)$ and $h(r)=h_\beta(r)$
for the non-Schwarzschild Anti-de Sitter case.} \label{fig1}
\end{figure}

For the case of asymptotically de Anti-de Sitter black holes ($\Lambda<0$) there is not cosmological horizon.
Thus, the conditions used previously that $f$ and $h$ vanish at cosmological horizon is no more valid and
this implies that one has to adopt the boundary condition at infinity as the criterion of the shooting method.

\begin{figure}[h]
\includegraphics[width=5.5cm]{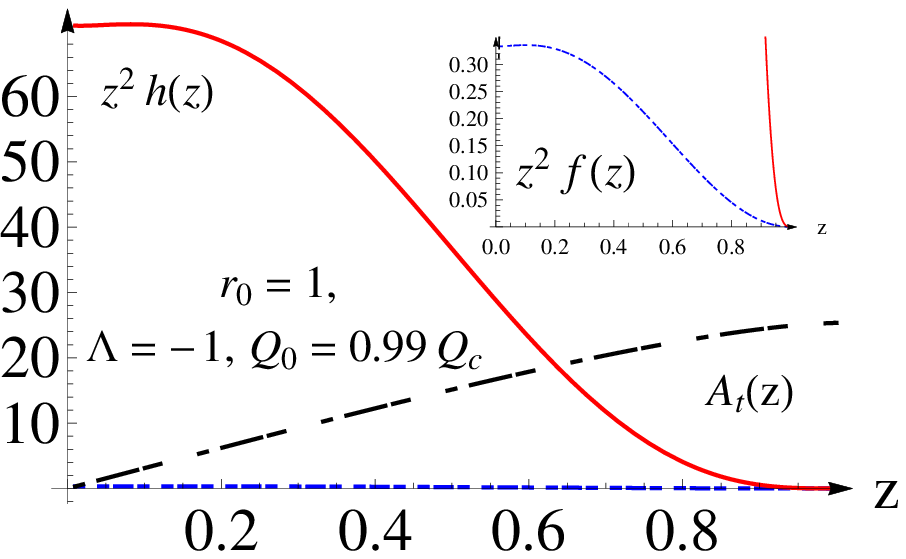}\includegraphics[width=5.5cm]{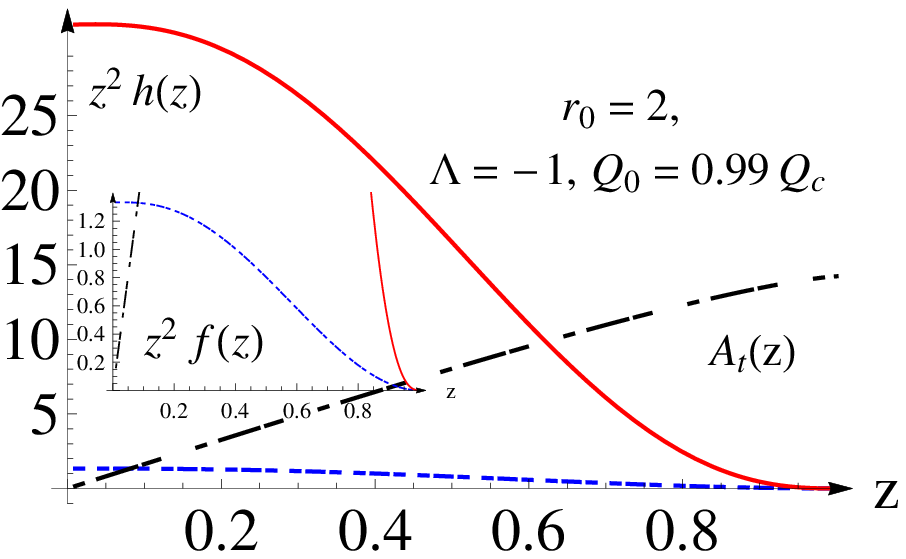}
\includegraphics[width=5.5cm]{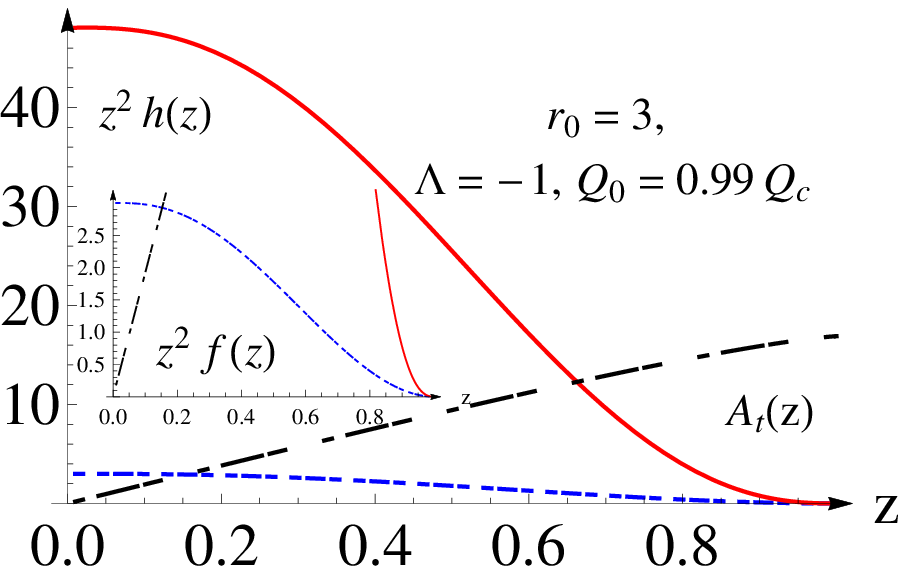}
\includegraphics[width=5.5cm]{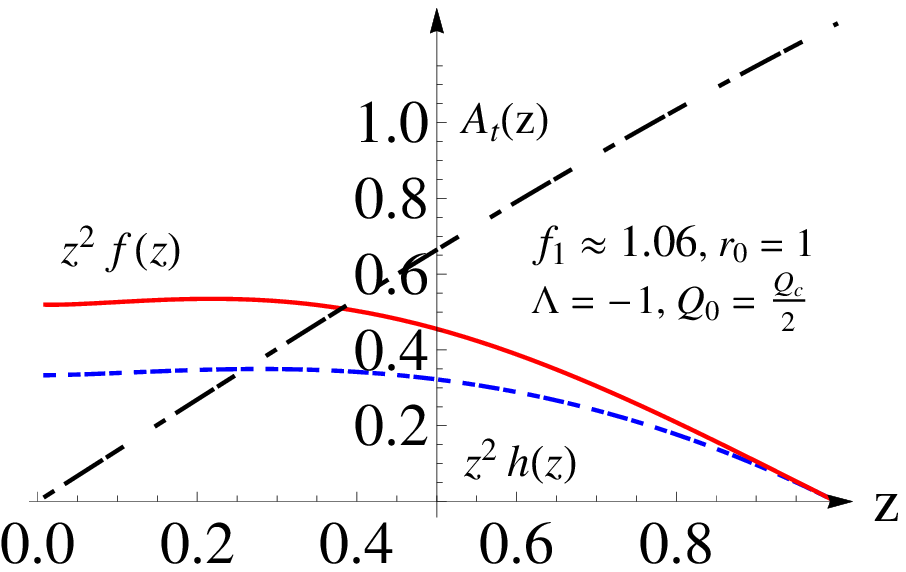}\includegraphics[width=5.5cm]{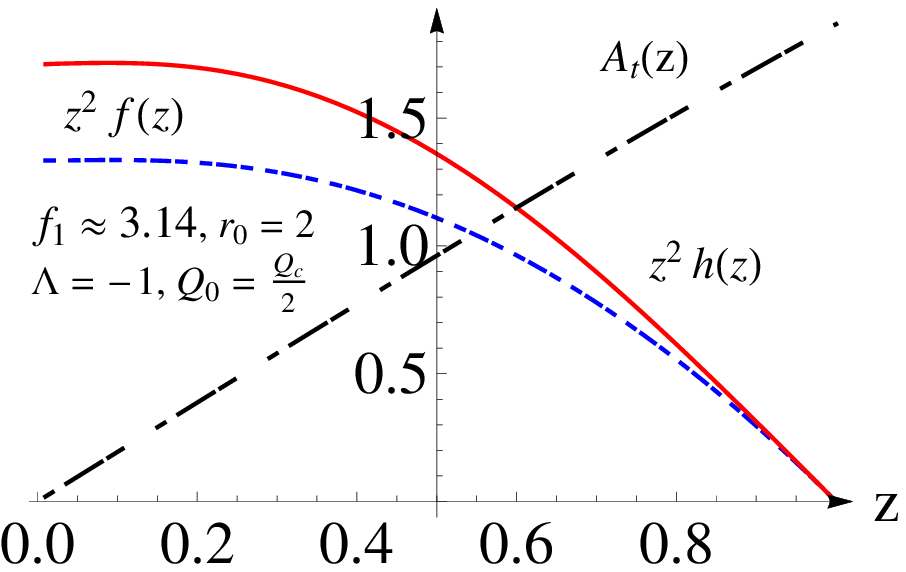}
\includegraphics[width=5.5cm]{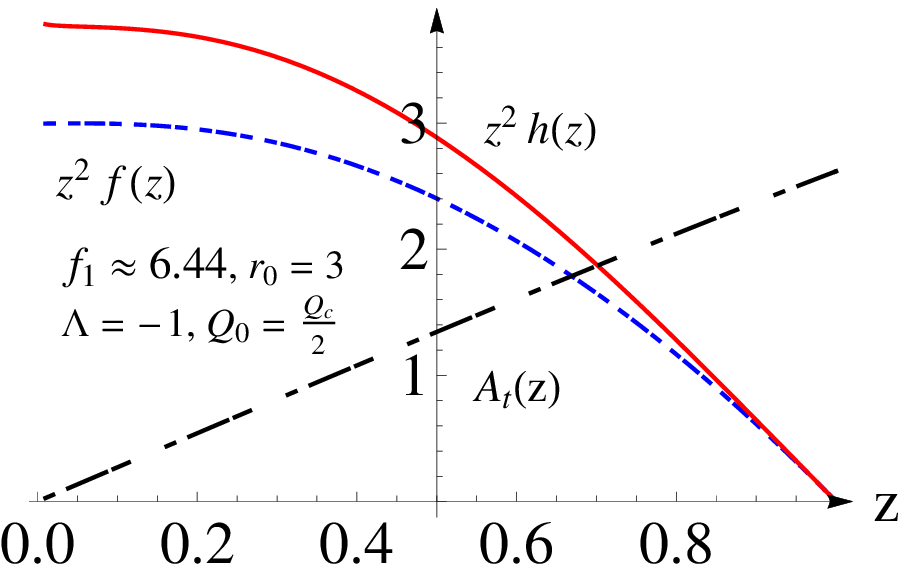}
\caption{ Anti-de Sitter solutions belonging to group I. We display
$f(r)$, $h(r)$ and $A_t(r)$ with $r_0=1,2,3$ and $Q_0=0.99Q_c,\frac{Q_c}{2}$, where $Q_c$ is the charge of the
extreme Anti-de Sitter black hole.} \label{fig3}
\end{figure}

\begin{figure}[h]
\includegraphics[width=6cm]{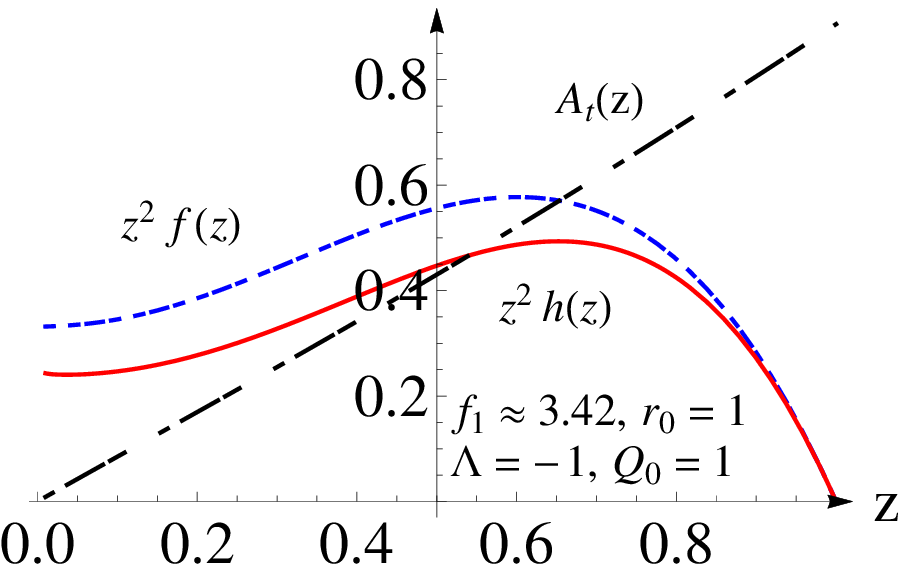}\includegraphics[width=6cm]{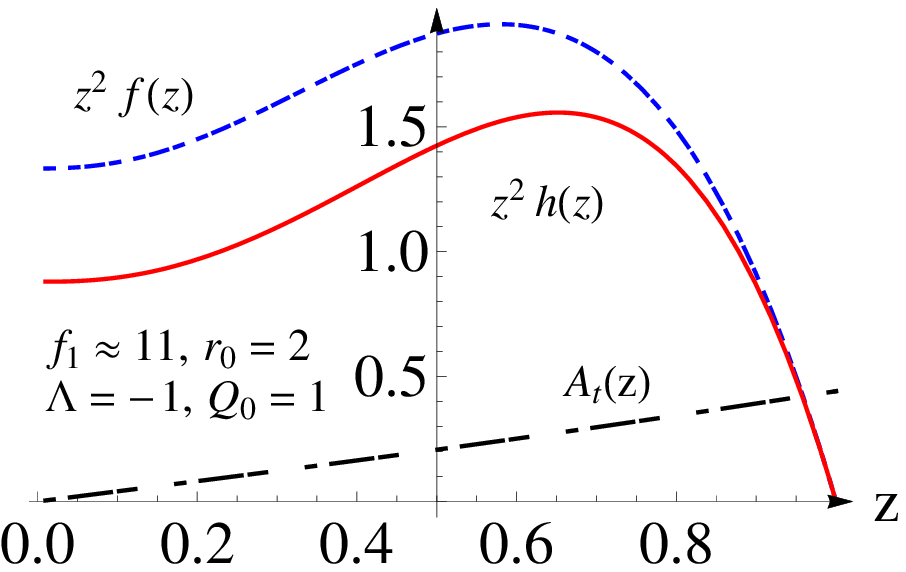}\includegraphics[width=6cm]{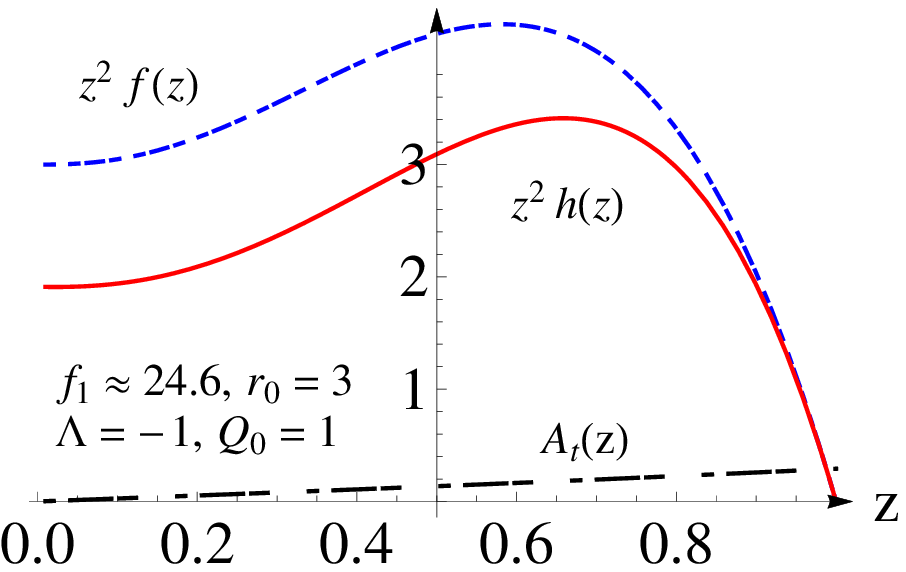}
\includegraphics[width=6cm]{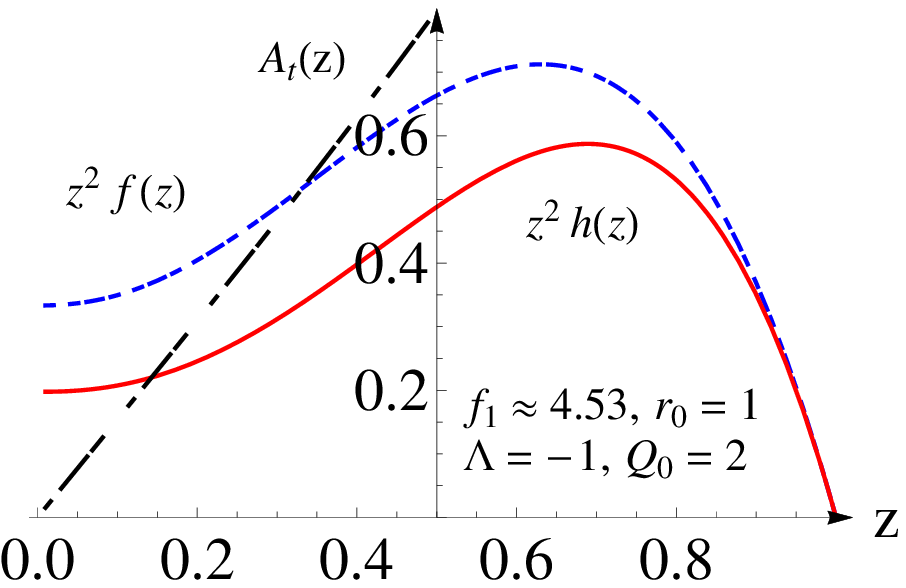}\includegraphics[width=6cm]{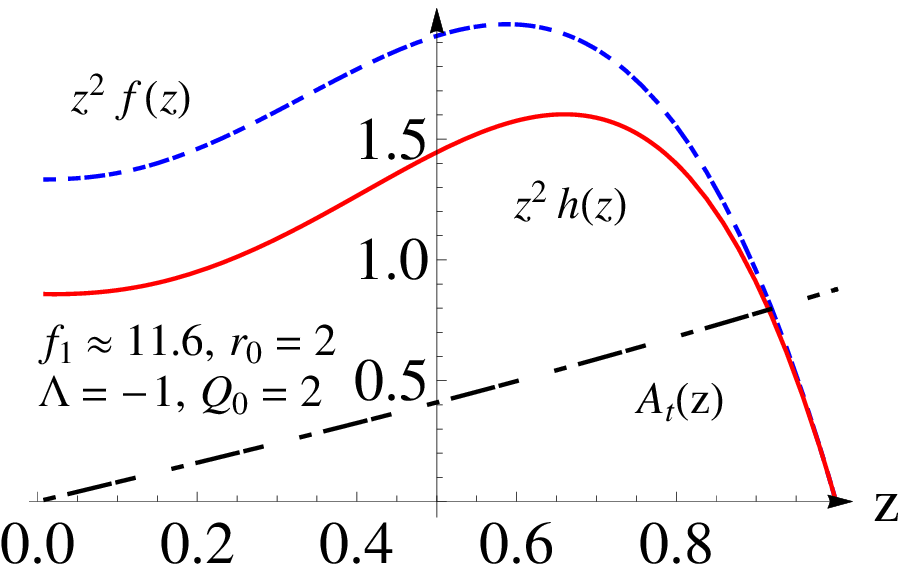}\includegraphics[width=6cm]{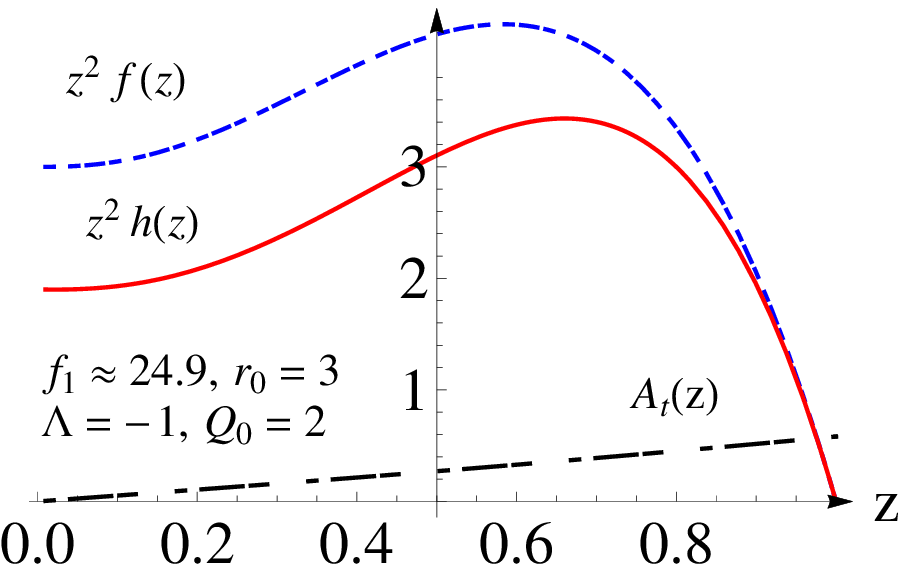}
\includegraphics[width=6cm]{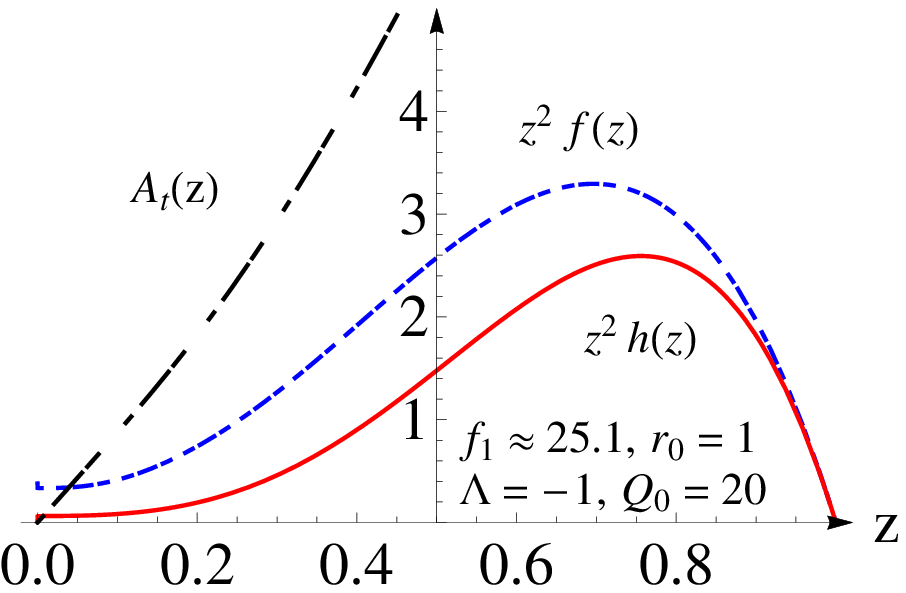}\includegraphics[width=6cm]{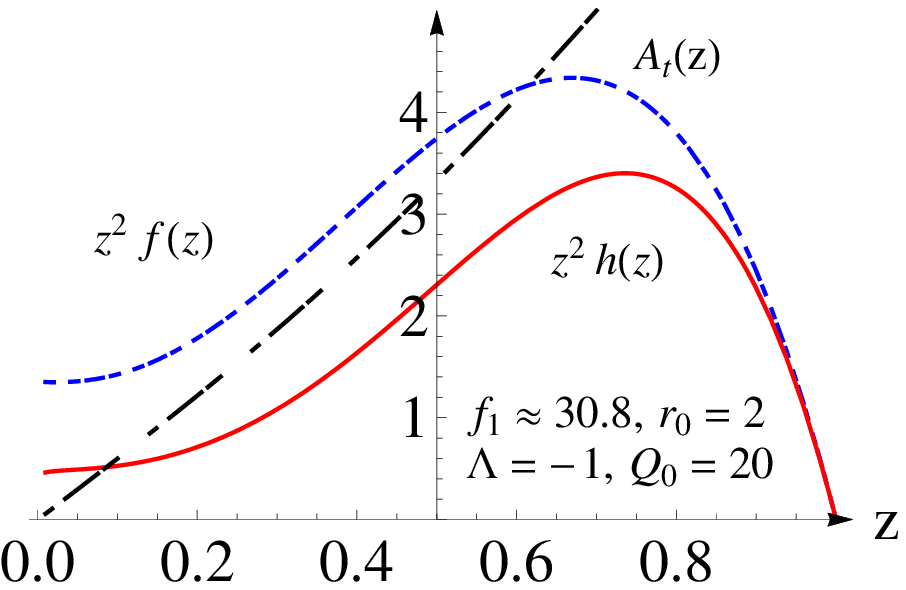}\includegraphics[width=6cm]{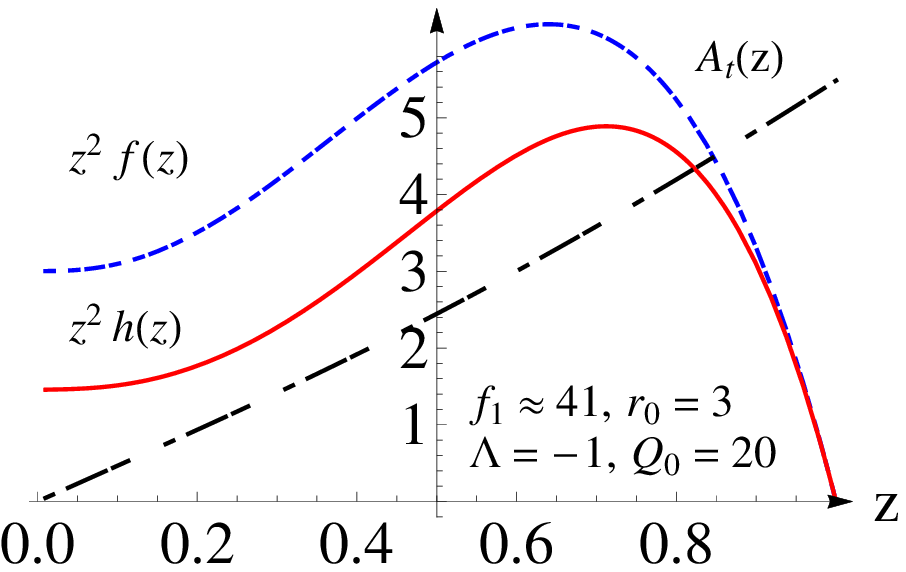}
\caption{Numerical Anti-de Sitter solutions group II of $f(r)$, $h(r)$ and $A_t(r)$ with $r_0=1,2,3$ and $Q_0=1,2,20$.
} \label{fig4}
\end{figure}

Since it is known that $f$ and $h$ approach asymptotically the form $r^2$ as $r\rightarrow\infty$, we introduce
the coordinate transformation $z=\frac{r_0}{r}$, so that the function of $f$ and $h$ become $f=f(z)$ and $h=h(z)$
where $0 \le z \le 1$. When $z=1$, the corresponding radial coordinate is the event horizon $r_0$, and $z=0$
corresponds to $r\rightarrow\infty$.
Therefore, with this transformation it is possible to apply the boundary conditions at finite values of $z=0$ and $z=1$.
In our calculations, the condition that $f$ and $h$ are proportional to $z^{-2}$ at $z=0$ is used as the criterion
for the shooting method: at $z=0$, the functions $z^2f(z)$ and $z^2h(z)$ should be  positive  and their
derivatives with respect to $z$ should vanish.
We show the resulting Anti-de Sitter black hole in the Figs.(\ref{fig1},\ref{fig3},\ref{fig4}).

Without loss of generality, we set $\Lambda=-1$. Similar to case of de Sitter spacetime, we find that the field
equations possess two groups of solutions, which reduce to Schwarzschild Anti-de Sitter and non-Schwarzschild
Anti-de Sitter black holes respectively as $Q_0 \rightarrow 0$.
We show the uncharged Anti-de Sitter solutions in figure \ref{fig1}. It is observed that the difference between
Schwarzschild Anti-de Sitter and non-Schwarzschild anti-de Sitter black holes increases as $r_0$ increases.

Again, the Reissner-Nordstr\"om Anti-de Sitter metric is not a solution of the field equations, while the group I
solutions reduce to Schwarzschild Anti-de Sitter black hole as $Q_0 \rightarrow 0$. Surprisingly, for this theory
of gravity, an ``extreme" Anti-de Sitter black hole solution is possible.

The critical charge $Q_c^{AdS}$ of the ``extreme" Anti-de Sitter black hole for group I is determined by
\bqn
\label{eq8a}
Q_c^{AdS}=\frac{r_0}{\sqrt{3\kappa}}\sqrt{(r_0^2|\Lambda|+1)(8\alpha|\Lambda|+3)}\quad .
 \eqn
However, for the group II solutions, no maximal charge is found. We show numerical solutions for group I and
II in figure \ref{fig3} and figure \ref{fig4} respectively.

In this letter, we investigated (Anti-) de Sitter electrically charged static black hole solutions of
Einstein-Weyl gravity. Two groups of solutions were found for the theory.
Group I is characterized by an upper limit for the black hole's charge corresponding to the extreme (Anti-) de
Sitter black hole, while group II does not possess any maximal charge.
These conclusions are consistent with the results found in \cite{LPPS,LPFA}.
Recently, the quasinormal modes of non-Schwarzschild solutions have been studied in \cite{Cai}
and it was shown that the black hole spacetime is stable.
It is therefore interesting to investigate the quasinormal modes and stability of the charged static solutions of
Einstein-Weyl theory in (Anti-) de Sitter spacetime. In particular {the latter may} have important
implications on the AdS/CFT correspondence since the non-Schwarzschild AdS black hole could be a new thermal
state in CFT.

\section*{\bf Acknowledgements}

This work is supported in part by Brazilian funding agencies FAPESP, FAPEMIG, CNPq, CAPES, and by Chinese
funding agencies NNSFC under contract No.11573022 and 11375279.

\onecolumngrid

\end{document}